\documentclass[fleqn,10pt]{wlscirep}

\usepackage[utf8]{inputenc}
\usepackage[T1]{fontenc}
\usepackage{lineno}
\linenumbers
\usepackage{color,soul}
\usepackage{upgreek}
\usepackage{mathastext}
\usepackage{setspace}
\usepackage{eqnarray,amsmath}
\usepackage{etoolbox}
\usepackage{lmodern}
\usepackage{array}  
\usepackage{xcolor}

\usepackage{geometry}  
\usepackage{pdflscape}  
\usepackage{longtable}  
\title{\centering \textcolor{black}{Training of  Physical Neural Networks}}

\author[1]{Ali~Momeni}
\author[2]{Babak~Rahmani}
\author[3]{Benjamin~Scellier}
\author[4]{Logan~G.~Wright}
\author[5]{Peter~L.~McMahon}
\author[6]{Clara~C.~Wanjura}
\author[7]{Yuhang~Li}
\author[8]{Anas~Skalli}
\author[9]{Natalia~G.~Berloff}
\author[5,10]{Tatsuhiro~Onodera}
\author[11]{Ilker~Oguz}
\author[12]{Francesco~Morichetti} 
\author[13]{Philipp~del~Hougne}
\author[14]{Manuel~Le~Gallo}
\author[14]{Abu~Sebastian}
\author[15,16]{Azalia~Mirhoseini}
\author[2]{Cheng~Zhang}
\author[17]{Danijela~Marković}
\author[8]{Daniel~Brunner}
\author[11]{Christophe~Moser}
\author[18]{Sylvain~Gigan}
\author[6]{Florian~Marquardt}
\author[7]{Aydogan~Ozcan}
\author[19]{Julie~Grollier}
\author[20]{Andrea~J.~Liu}
\author[21]{Demetri~Psaltis}
\author[22,23]{Andrea~Alù}
\author[1,*]{Romain~Fleury}

\affil[1]{Laboratory of Wave Engineering, School of Electrical Engineering, Swiss Federal Institute of Technology in Lausanne (EPFL), Lausanne, Switzerland}
\affil[2]{Microsoft Research, 198 Cambridge Science Park, CB4 0AB Cambridge, UK}
\affil[3]{Rain AI, San Francisco, USA}
\affil[4]{Department of Applied Physics, Yale University, CT, USA}
\affil[5]{School of Applied and Engineering Physics, Cornell University, Ithaca, New York 14853, USA}
\affil[6]{Max Planck Institute for the Science of Light, Staudtstraße 2, 91058 Erlangen, Germany}
\affil[7]{Department of Electrical and Computer Engineering, University of California, Los Angeles, CA 90095, USA}
\affil[8]{FEMTO-ST Institute/Optics Department, CNRS \& University Bourgogne Franche-Comté, Besançon Cedex 25030, France}
\affil[9]{Department of Applied Mathematics and Theoretical Physics, University of Cambridge, Cambridge, UK}
\affil[10]{NTT Physics and Informatics Laboratories, NTT Research, Inc., Sunnyvale, USA}
\affil[11]{Applied Photonics Devices (LAPD), École Polytechnique Fédérale de Lausanne (EPFL), Lausanne, Switzerland}
\affil[12]{Dipartimento di Elettronica, Informazione e Bioingegneria, Politecnico di Milano, Milan, Italy}
\affil[13]{Univ Rennes, CNRS, IETR – UMR 6164, F-35000 Rennes, France}
\affil[14]{IBM Research Europe– Zurich, 8803 Rüschlikon, Switzerland }
\affil[15]{Department of Computer Science, Stanford University, USA}
\affil[16]{Google DeepMind, 1600 Amphitheatre Parkway Mountain View, CA 94043.}
\affil[17]{Unité Mixte de Physique CNRS/Thales, CNRS, Thales, Université Paris-Saclay, Palaiseau, France}
\affil[18]{Laboratoire Kastler Brossel, Sorbonne Université, École Normale Supérieure, Collège de France, CNRS UMR 8552, Paris, France}
\affil[19]{Laboratoire Albert Fert, CNRS, Thales, UniversitéParis-Saclay, Palaiseau91767, France}
\affil[20]{Department of Physics and Astronomy, University of Pennsylvania, Philadelphia, PA, 19104, USA}
\affil[21]{Optics Laboratory (LO), École Polytechnique Fédérale de Lausanne (EPFL), Lausanne, Switzerland}
\affil[22]{Photonics Initiative, Advanced Science Research Center, City University of New York, New York, NY, 10031, USA}
\affil[23]{Physics Program, Graduate Center, City University of New York, New York, NY, 10016, USA}

\affil[*]{E-mail: romain.fleury@epfl.ch}

\begin{abstract}
Physical neural networks (PNNs) are a class of neural-like networks that leverage the properties of physical systems to perform computation.
While PNNs are so far a niche research area with small-scale laboratory demonstrations, they are arguably one of the most underappreciated important opportunities in modern artificial intelligence (AI). Could we train AI models 1000x larger than current ones? Could we do this and also have them perform inference locally and privately on edge devices, such as smartphones or sensors? 

Research over the past few years has shown that the answer to all these questions is likely ``textit{yes, with enough research}'': PNNs could one day radically change what is possible and practical for AI systems. To do this will however require rethinking both how AI models work, and how they are trained -- primarily by considering the problems through the constraints of the underlying hardware physics. To train PNNs at large scale, many methods including backpropagation-based and backpropagation-free approaches are now being explored. These methods have various trade-offs, and so far no method has been shown to scale to the same scale and performance as the backpropagation algorithm widely used in deep learning today. However, this is rapidly changing, and a diverse ecosystem of training techniques provides clues for how PNNs may one day be utilized to create both more efficient realizations of current-scale AI models, and to enable unprecedented-scale models. 
\end{abstract}

\begin{document}
\nolinenumbers

\maketitle

\section*{Introduction}

In recent years, artificial intelligence (AI) has profoundly influenced our daily lives through tools such as personal assistant chatbots, and has been utilized in various scientific fields such as healthcare, weather prediction, and material design to tackle some of the world's most challenging questions. Recent advancements in AI systems have been powered by the digital, silicon-based computing power of Graphics Processing Units (GPUs) as well as the unprecedented abundance of data. AI systems continue to evolve at an accelerated pace. With a clear trend toward increasingly larger models, the reliance on traditional digital GPUs is becoming untenable. The primary concerns are the high energy consumption (number of operations per second per Watt), low throughput (number of samples processed per second), and high latency caused by the separation of the memory and processing unit during the training and inference phases of AI systems. Given the widening gap between, on the one hand, the rapid increase in floating-point computation required for AI training and the slower improvements in computing hardware traditionally predicted by Moore's Law, and on the other hand the low data transfer rates between the memory modules and the computational cores, there has been renewed interest in alternative computing platforms, such as optical, photonics, and analog electronics. We collectively refer to these unconventional computing platforms as analog physical neural networks (PNNs).
Previous reviews have concentrated on the domain-specific technological advancement, applications and inference capabilities of PNNs majorly in optics \cite{wetzstein2020inference} and electronics \cite{sebastian2020memory}, among other media. In this article, we aim to explore the development of these PNNs from a training perspective, as broadly as possible, from the ground up and agnostic to the domain. We review methods that employ backpropagation---the learning mechanism most commonly used for digital-electronic neural networks. Additionally, we investigate approaches that minimise digital-electronic computation and leverage the inherent dynamics of the system to learn parameters of the analog system. Finally, we examine local training algorithms that minimize some local objectives which might be easier to implement without digital-electronic processing. Finally, we discuss the application of analog computers in handling larger models and the strategies for efficient training.

\begin{figure}
    \centering
    \includegraphics[width=1\textwidth]{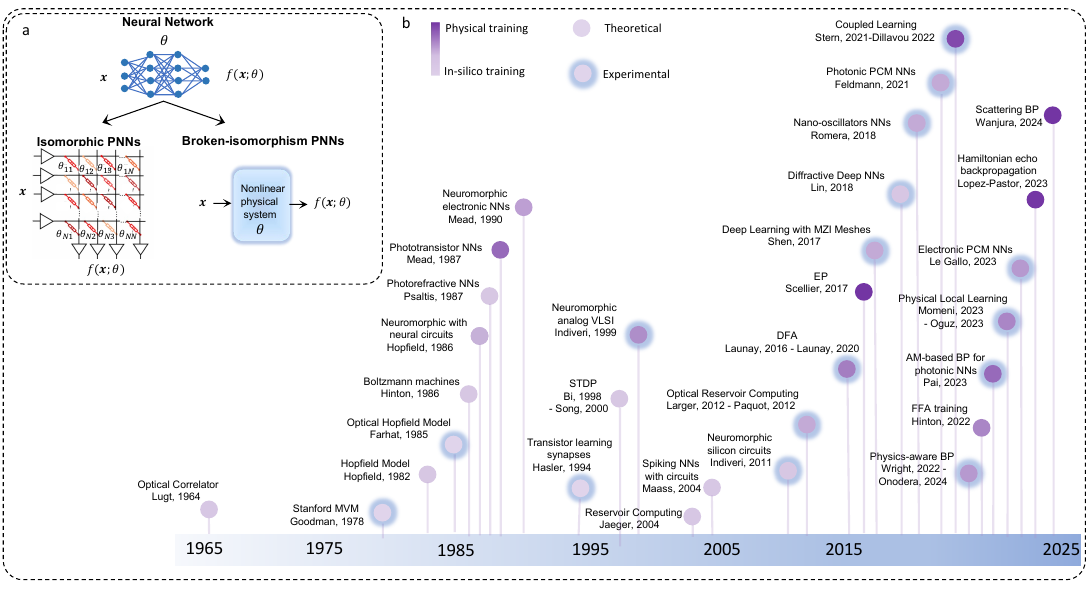}
    \caption{\textbf{a Physical Neural Networks (PNNs)}, processing input data $\vec{x}$ using trainable parameters $\vec{\theta}$. PNNs can be constructed to realize computations isomorphic to those commonly found in artificial neural networks, such as matrix-vector multiplications, or can sacrifice isomorphism for potential speed/energy advantages, where the physical system is left to perform the computation it most naturally performs. \textbf{b Timeline of training methods for PNNs.} The corresponding references of selected key milestones and publications from left to right:  \cite{lugt1964signal,goodman1984optical,hopfield1982neural,farhat1985optical,hinton1986learning,hopfield1986computing,psaltis1988adaptive,mead1987neural,mead1990neuromorphic,hasler1994single,bi1998synaptic,song2000competitive,indiveri1999neuromorphic,jaeger2004harnessing,maass2004computational,indiveri2011neuromorphic,larger2012photonic,paquot2012optoelectronic,nokland2016direct,launay2020hardware,scellier2017equilibrium,shen2017deep,lin2018all,romera2018vowel,feldmann2021parallel,stern2021supervised,dillavou2022demonstration,wright2022deep,onodera2024scaling,hinton2022forward,pai2023experimentally,momeni2023backpropagation,oguz2023forward,le202364,lopez2023self,Wanjura2023fully}. }
    \label{fig:history}
\end{figure}

\section*{Historical Overview of Analog Computing and PNNs}

Artificial neural networks (ANNs) were originally used to model biological neural networks starting in the 1930s \cite{mcculloch1943logical}. The field gained momentum with the invention of the perceptron by Warren McCulloch and Walter Pitts in 1943 \cite{mcculloch1943logical}, followed by Frank Rosenblatt’s hardware implementation in 1957 \cite{rosenblatt1958perceptron}. These developments marked a shift towards using ANNs in machine learning,  gradually deviating from their biological origins.

A natural evolution from the perceptron is the adaptive linear neuron classifier, or ADALINE. A key difference from the perceptron is that the linear activation function is used to train the weights, while the step function is only used during the inference phase\cite{denker1986neural}. 
Hebbian Learning Rule, also known as the Hebb Rule, was proposed by Donald O. Hebb \cite{hebb2005organization}. Hebb's rule can be described as a method of determining how to adjust the weights of a model. According to this principle, the weight between two neurons increases if the two neurons activate simultaneously and decreases if they activate separately.
Spike Timing-Dependent Plasticity (STDP) \cite{bi1998synaptic, song2000competitive} is a temporally asymmetric form of the Hebb Rule that is triggered by tight temporal correlations between the pre- and postsynaptic neurons' spikes. This type of local and event-based learning requires no extra energy for non-local transmission, as is needed during the training of ANNs. Based on this principle, spiking neural networks (SNNs), which have been proposed and implemented on neuromorphic platforms, may be more energy-efficient than ANNs \cite{hasler1994single,maass2004computational,indiveri1999neuromorphic,indiveri2011neuromorphic}.
Another milestone was the associative model proposed by Hopfield \cite{hopfield1982neural}. Hopfield's approach illustrates the method of collecting and retrieving memories based on the system's attributed energy. The memories correspond to the minima of the energy attributed to the system.
In the early 1980s, two other promising learning procedures for deep neural networks were proposed. One was Boltzmann Machines \cite{hinton1986learning} which performed unsupervised contrastive learning. Another was backpropagation \cite{rumelhart1986learning} which is now a widely used learning algorithm for training deep neural networks.

Despite the rapid advancement of artificial neural networks (ANNs) in digital processors, there has now been a renewed interest in implementing these networks in analog systems (see Figure \ref{fig:history}).
This idea was initially pursued in optical systems, such as holographic gratings \cite{psaltis1990holography}, and in electronic systems with bipolar transistors \cite{mead1987neural} as well as memristive-crossbar neural networks \cite{hubbard1986electronic}. Reservoir computing (RC) and Extreme Learning Machines (ELM) are two other historically important computational frameworks widely implemented in various fields \cite{tanaka2019recent}.
These frameworks treat PNNs as a non-trainable black-box that non-linearly transforms input data. The PNN may have rich recurrent dynamics and thus can have "memory" for time-dependent tasks (RC) or it may simply be a pure feed-forward system suitable for static input data (ELM).
RC was first introduced under the names "echo-state network" \cite{jaeger2001echo} and "liquid state machine" \cite{maass2002real}. 
%This approach simplifies the use of recurrent neural networks (RNNs) by eliminating the need for intensive back-propagation training. 
In RC, input data is transformed through a high-dimensional network of interconnected nonlinear nodes—the reservoir. Usually, the RC output consists of a linear combination of node responses, and importantly, only the weights of this final output layer are trained, making it resource-efficient and reducing complexity by maintaining input and internal coupling weights constant. 
%The Extreme Learning Machine (ELM) is highly similar to RC, where the recurrent network is replaced with a feed-forward architecture. 
More importantly, because only a simple output layer is trained, RC and ELMs can be thought of as the simplest trainable framework for  PNNs. Indeed, the reservoir can be any non-linear, sufficiently high-dimensional physical system. Neurons are embedded in the physical system’s degrees of freedom, giving RC unprecedented flexibility regarding physical implementation. Internal coupling is usually realized through the physical system’s inherent interactions, and input weights correspond to coupling the system to an external signal that drives the system’s various dimensions. Implementing the output layer is crucial and can be realized either in hardware i.e. online with physical devices or in software i.e. offline via digital weights that combine previously measured neuron states. RC and ELMs have been implemented on a wide variety of physical substrates ranging from electronics \cite{appeltant2011information} and spintronics \cite{markovic2019reservoir}, optics for using a time-multiplexed approach \cite{brunner2013parallel}, frequency \cite{lupo2023deep,momeni2022electromagnetic} and spatial \cite{skalli2022computational,rafayelyan2020large} multiplexing, to exciton-polariton condensates \cite{ballarini2020polaritonic} and mechanical substrates \cite{nakajima2013soft}. Recently, there have also been efforts for quantum RC proposals and implementations to leverage the exponential scaling of the Hilbert space as a high-dimensional mapping for information processing \cite{fujii2017harnessing,markovic2020quantum,ghosh2021quantum,govia2021quantum,mujal2021opportunities,yasuda2023quantum,hu2023overcoming,senanian2023microwave}.
Most physical implementations of RC leverage software weights for their high flexibility. Indeed, they provide single-shot learning via a simple matrix inversion, the ability to leverage existing sensory data, particularly in the field of robotics and edge computing \cite{nakajima2013soft,nakajima2020physical} and easier integration with follow-up models that would leverage RC as an efficient physical pre-processor \cite{teugin2021scalable}. In contrast, the implementation of hardware weights to fully exploit capabilities of the physical system such as inference speed and latency \cite{skalli2022computational}, hardware weights also offer potentially substantial energy savings and scaling as an in-memory alternative to their software counterpart and require hardware-compatible training strategies\cite{oguz_programming_2024,syed2023beyond}. 
Moreover, depending on the physical substrate, additional challenges may arise, particularly when it comes to memory-dependent tasks. These tasks imply that input information should drive the physical system at a rate in accordance with its own intrinsic timescales, which can be challenging for ultra-fast systems and entails fast modulation bandwidths.
\\
\\
\\
\fbox{\begin{minipage}{49em}

\section*{Box1: PNNs} 
\footnotesize
Physical Neural Networks (PNNs) are defined as physical systems involving weights ($\theta$) that can be adjusted in order to learn and perform a desired computing task. PNNs resemble neural networks, however at least part of the system is analog rather than digital, meaning that part or all the input/output data is encoded continuously in a physical parameter, and the weights can also be physical, with the ultimate goal of surpassing digital hardware in performance or efficiency. PNNs are divided into two categories, depending on whether or not they mimic digital neural networks: isomorphic PNNs and broken-isomorphism PNNs, respectively. Isomorphic PNNs perform mathematical transformations by designing hardware for strict, operation-by-operation mathematical isomorphism, such as memristor crossbars for performing matrix-vector multiplications (see Fig. 1a). In contrast, broken-isomorphism PNNs break mathematical isomorphism to directly train the hardware's physical transformations \cite{wright2022deep}. One complication with broken-isomorphism PNNs is that it is often unknown what features are required for universal computation or universal function approximation. The notion of trainable broken-isomorphism PNNs emerged, in part, from untrained physical systems being used for machine learning: physical reservoir computing \cite{tanaka2019recent}. There were also several theoretical proposals \cite{hughes2019wave,khoram2019nanophotonic,nakajima2021neural} of broken-isomorphism PNNs prior to the general framework and experimental demonstrations presented in Ref. \cite{wright2022deep}. Broken-isomorphism PNNs could potentially perform certain computations much more efficiently than digital methods, leading to a path for more scalable, energy-efficient, and faster machine learning.
%Analog PNNs inevitably have noise, which can harm their inference accuracy if it is not considered when choosing how to operate the PNN \cite{yang2022tolerating}. 
Minimizing the power consumption of a PNN will generally result in a reduction of the signal-to-noise ratio in the PNN, even more strongly motivating the need to deal with noise. One strategy is to treat the stochasticity of the physical system as a resource that can be harnessed in machine-learning applications that are fundamentally probabilistic, such as generative models \cite{wu2022harnessing,coles2023thermodynamic} or Bayesian models \cite{bonnet2023bringing}. Alternatively, one can train the PNN to perform deterministic inference tasks in a way that is resilient to noise \cite{olin2021stochasticity},  even in the limit where the signal-to-noise ratio is low ($\sim1$) and quantum noise dominates\cite{ma2023quantum}. \label{box1}
\end{minipage}}

%\subsection*{Extreme Learning Machines and Reservoir Computing}

\section*{Training Techniques of PNNs}
\subsection*{In-Silico Training}

In-silico methods for training PNNs involve digitally emulating and optimizing the physical degrees of freedom ($\theta$) of the hardware, followed by deploying this optimized physical architecture \cite{lin2018all, hughes2019wave,yu2020neuromorphic, sunada2021physical, lebleici2017neuromorphic, englund2023vcsel, wetzstein2018hybrid, eleftheriou2020accurate, narayana2023hardware, sebastian2023core, rahmani2020actor} during the inference phase, as summarized in box2. These methods employ physics-based forward models and/or digital neural networks to create digital twins of PNNs within a computer environment, which are optimized for a specific task. After this optimization process, the resulting PNN hardware is deployed for the analog processing of new data.

In-silico training methods enable rapid exploration, validation and testing of various PNN architectures, helping to improve the accuracy and functionality of PNNs before they are physically constructed. This approach is notably faster and cost-effective, eliminating the need to set up and optimize expensive and time-consuming physical systems for each iteration of the design; this also allows scalability, where the PNN architecture can be adjusted and expanded as needed. When the hardware to be deployed does not have significant nonlinearities or device defects, which would require more complex and specific information to be included during training, an approximate digital model with Gaussian noise injected during the forward pass can be sufficient in many cases to obtain accurate inference results on hardware \cite{eleftheriou2020accurate, sebastian2023core}. In-silico training also ensures scientific reproducibility and transparency, which is an important advantage compared to some of the delicate and expensive in-situ learning systems that are harder to replicate. Lastly, in-silico methods enable the exploration of theoretical/Gedanken models and PNN structures that are beyond the capabilities of current technological constraints (e.g., fabrication resolution/precision, material properties, optical nonlinearities, losses, etc.) and help us analyze the effects of various factors in a controlled environment.

However, in-silico training comes with its own set of limitations, as emphasized in Box~2. One of the challenges is related to the digital forward model of the physical architecture. The physical/hardware complexity of the PNN architecture might pose a challenge in identifying appropriate analytical and/or numerical models to digitally represent it accurately, which can limit the effectiveness of in-silico learning. Digital forward models might fail to encompass all physical phenomena in the actual PNN hardware, such as detection noise, misalignments, fabrication and material imperfections \cite{momeni2023backpropagation, wright2022deep}, among other experimental factors; this forms a challenge for the accurate deployment of these trained PNNs at large scale, covering many devices. The computational demands of these forward model simulations form another potential hurdle. The process of discretizing the continuous physical world requires finer grids for improved accuracy, which can lead to exponential increases in computational requirements with the scaling of the physical size of the PNN and its input/output channels \cite{shimobaba2009band, matusik2021towards}.
Another limitation is that this method can only be as fast and efficient as digital computers. It will typically be much less efficient than training conventional digital neural networks since modeling PNN hardware will come with computational overhead. 

%Despite these challenges, in-silico training is still an indispensable tool in developing PNNs, providing a foundational platform for designing and transitioning virtual models to tangible PNN applications.

\subsection*{Physics-aware BP Training}
Transitioning from gradient-free in-situ optimization to the favorable scaling of hybrid in-situ--in-silico backpropagation algorithms \cite{Adhikari2012, Cramer2022, Spall2022} has been a critical step for the emerging field of PNN training. Physics-aware Training \cite{wright2022deep} (PAT) crystallizes the notion that this can be reliably done for any physical system with an approximate predictive model. In this algorithm, the physical system performs the forward pass, and the backward pass is performed by differentiating the digital model. Its key feature, a mismatched forward and backward pass, which is shared with many training algorithms \cite{Lillicrap2016,hubara2018quantized,launay2020direct, Adhikari2012, Cramer2022, Spall2022}, requires only that the digital model produce an estimated gradient that is approximately aligned with the true gradient \cite{Lillicrap2016}. This condition, which is much less stringent than requiring a perfect digital model, allows PAT in many cases to be a drop-in replacement for in-silico training, with many of the benefits of in-situ training algorithms. Demonstrating its versatility, PAT has successfully trained PNNs across various domains including optical, mechanical, and electronic systems \cite{wright2022deep}.

Physics-aware training is a hybrid of in-situ and in-silico methods, inheriting strengths and weaknesses from both. Owing to its in-situ component, PAT mitigates the effect of experimental noise and mismatches between the experiment and the digital model. Meanwhile, its in-silico nature enables accurate training with time scaling similar to backpropagation. However, training may be slow if the physical system’s parameters can only be updated slowly. Although constructing a digital model for PAT is less demanding than for pure in-silico methods, it still poses challenges for complex, large-scale PNNs. Given the difficulties in constructing accurate physics-based models (see In-Silico Training Section), the emerging field of physics-informed machine learning, which integrates data-driven methods with physical principles, offers a promising solution \cite{Brunton2022,Karniadakis2021}. In this vein, an optical wave simulation model with data-driven fine-tuning was used to perform PAT on a complex PNN with 10,000 parameters in Ref.~\cite{onodera2024scaling}.
\\
\\
\\
\fbox{\begin{minipage}{48em}

\section*{Box2: Inference and Training Processes in PNNs} \label{box2}
\footnotesize
PNNs embody the cutting-edge convergence of physical hardware, material science and artificial intelligence, operating in a two-step process: training and inference. In the training phase, a PNN learns from data related to a specific inference task, adjusting its physical degrees of freedom ($\theta$) based on the corresponding feedback to minimize the difference between its outputs and ground truth. Central to the training of PNNs is the use of error backpropagation, a method used to calculate the gradients of a desired loss function with respect to network weights. This process can be encapsulated by modeling the PNN as a parameterized function $f_\text{physical}$ that relates the input ($\mathbf{x}$) and the physical parameters ($\mathbf{\theta}$) of the system with the output ($\mathbf{y}$). Given that, for many physical models, there currently exist no straightforward method to extract the weight gradients, a digital twin ($f_\text{digital}$) is often used to approximate $f_\text{physical}$. Stochastic gradient descent, commonly used in deep learning, can be applied to optimize PNNs through the following steps:

	- Forward Model Execution:  $\mathbf{y}=f_* (\mathbf{x};\mathbf{\theta}) $

	- Loss Computation: $L=l(\mathbf{y},\mathbf{y_\text{target}})$

	- Error Backpropagation and Parameter Update:  $\mathbf{g_\theta}=\frac{\partial L }{\partial \theta}= \frac{\partial L}{\partial y} \frac{\partial y}{\partial \theta} ; \mathbf{\theta} \rightarrow \mathbf{\theta}- \mu \mathbf{g_\theta}$

where $f_*$ denotes either $f_\text{digital}$ or $f_\text{physical}$, corresponding to in-silico or in-situ training methods, respectively. $l$ is the loss function, $\mu$ is the learning rate and $\mathbf{y_\text{target}}$ is the target output (i.e., the ground truth). For forward model execution ($f_*$), both $f_\text{digital}$ and $f_\text{physical}$ can be used \cite{lin2018all,wright2022deep,hughes2019wave,spall2022hybrid}; however, for error backpropagation, $f_\text{digital}$ is more commonly employed due to the inherent challenges of applying traditional backpropagation to $f_\text{physical}$ \cite{ashtiani2022chip,liu2022programmable}. Despite being challenging, recent advancements have introduced in-situ learning methods that utilize error backpropagation-free training \cite{shen2017deep,momeni2023backpropagation,oguz2023forward} or backpropagation adaptations using the physical forward model of the PNN \cite{zhou2020situ,hughes2018training,pai2023experimentally}. Once the training is complete, unlike conventional neural networks that only operate digitally, an optimized PNN physically performs its inference on new data through $f_\text{physical}$ operating in the analog domain.
The implementation of PNNs introduces some unique challenges and trade-offs across the training and deployment (blind inference) phases. A key obstacle is the reliance on a digital twin—the mathematical representation of the PNN— which may not fully capture the complexities of the actual physical model, potentially overlooking various factors like fabrication imperfections, misalignments, and detection noise, among others. This might impact the inference accuracy of an in-silico trained PNN when it transitions from $f_\text{digital}$ to $f_\text{physical}$ in the deployment phase \cite{wright2022deep,mengu2020misalignment}. In-situ training of a PNN through $f_\text{physical}$ can circumvent some of these limitations. In either case, a PNN’s forward operation necessitates robust stability; for example, temporal variations in $f_\text{physical}$ due to mechanical/physical drifts or temperature fluctuations etc., would hurt both the training and inference phases regardless of which forward model ($f_*$) is used.
Arguably, this requirement for stability is one of the most significant challenges of PNN-based information processing and inference for real-world deployment, and it requires the marriage of advanced micro-/nano-fabrication methods along with material engineering and packaging for building resilience against external conditions. %Moreover, there exists a notable trade-off between the complexity of PNNs and their performance: larger PNNs, in general, improve performance but also increase physical complexity, alignment and stabilization requirements as well as costs. Balancing these factors—performance vs. complexity, cost, resource use, and stability—is a pivotal aspect of developing effective and efficient PNNs addressing real-world applications.
\end{minipage}}

\subsection*{Feedback Alignment}
The end-to-end BP faces notable challenges for in-hardware implementations due to gradient communication in the backward pass across all layers
\cite{momeni2023backpropagation,hinton2022forward,laydevant2023benefits}. BP uses the transpose of the weight matrix at each layer to back-propagate the error from the output to the input layer,  which requires extensive knowledge of the parameters of the NN, and requires that the same weights used in the forward pass are transposed and used in the backward pass, giving rise to "the weight transport problem". In contrast, both Feedback Alignment (FA) and Direct Feedback Alignment (DFA) were introduced as methods that allow training a NN without transferring weights from the forward pass to the backward pass for increased efficiency, usually at the cost of performance. Nevertheless, both methods still require the derivative of activation functions and the states of activation at each layer.  FA was first introduced in \cite{lillicrap2014random} and presents itself as a simple alternative for training NNs as it uses fixed, random feedback weights at each layer to propagate error signals from the output to earlier layers. This can significantly reduce the computational cost of the backward pass. In \cite{nokland2016direct}, an improved FA algorithm called direct feedback alignment (DFA) was introduced to address the main shortcomings of FA. DFA improves on FA by using fixed random feedback weight matrices to broadcast error signals directly to all layers simultaneously, thereby enabling the successful training of deeper networks. FA methods use fixed random projections to train NNs they are directly more suited to hardware implementations than traditional BP as fixed random projections are easier to implement in physical hardware  \cite{launay2020hardware}. 
Yet, despite its low complexity and apparent compatibility with physical implementation at first glance, physical realizations of DFA are scarce, as it still requires partial knowledge of the NN parameters and, in particular, the activation function of the NN. In \cite{nakajima2022physical}, under the name augmented DFA, DFA was extended to no longer require precise knowledge of the network's activation function and instead replaces it with an arbitrary function that still presents a correlation with said activation function. Augmented DFA was also used to train a neural network architecture utilizing a physical nonlinearity using a Mach-Zehnder modulator. Wang et al. \cite{wang2024asymmetrical} also proposed an asymmetrical estimator based on alignments for in-situ training of photonic neural networks.  Additionally, in the context of hardware acceleration for training NNs, a physically implemented DFA algorithm was implemented using high-dimensional random mappings in the optical domain to train digital NNs, using both fully connected and graph convolutional networks in \cite{launay2020hardware}, paving the way for the future use of DFA in training PNN.
Despite these advantages, the DFA suffers from accuracy degradation problem. This becomes more serious when the DFA is applied to convolutional and recurrent neural networks \cite{han2019efficient,refinetti2021align,han2020extension}. Also, this method is only compatible with certain PNNs, where it is possible to separate the nonlinear part and linear layer\cite{nakajima2022physical}.

\subsection*{Physical Local Learning}

Another solution to the stagnant gradient problem of end-to-end BP is local learning by eliminating gradient communication entirely. Each layer (or block) independently calculates and applies parameter updates using its own training signal. This setup ensures that no block remains idle, waiting for gradients from others, making it an optimal setting for distributed model optimization. Local objectives were initially used in early unsupervised methods for pre-training deep neural networks, such as the wake-sleep algorithm \cite{hinton1995wake}, Restricted Boltzmann Machine (RBM) \cite{salakhutdinov2007restricted}, and autoencoders (AE) \cite{vincent2010stacked}. The idea behind local training is twofold: one, to minimize a local loss that compresses the information; and two, to extract sufficient information from the input for the next layer/block. These compression and preparation for the next layer could be explained from an information bottleneck perspective \cite{Federici2020Learning}.  More recently, several variations on the local learning paradigm have been introduced, focusing on parallel training.  Löwe et al. \cite{lowe2019putting}  use a contrastive predictive loss to perform excellently in an unsupervised setting. Nøkland and Eidnes \cite{nokland2019training} and Ren et al. and Siddiqui et al. \cite{ren2023scaling,siddiqui2023blockwise} consider supervised and self-supervised local learning, respectively; and succeed in matching the accuracy of global learning on classification tasks with up to 100 classes. Some proposals introduced a coupling between subsequent blocks, allowing the gradient signal to flow between pairs of adjacent blocks \cite{xiong2020loco}. This strategy preserves many of the compute efficiency advantages of local optimization while recovering much of the task performance achieved by global optimization \cite{ gomez2022interlocking}. Local parallelism allows for fully asynchronous layer-wise parallelism with a minimal memory footprint. For instance, in reference \cite{lowe2019putting}, the architecture is divided into three independently trainable blocks, resulting in a 2.8 times reduction in GPU memory usage. Generally, the GPU memory will decrease approximately by a factor of $k$, where $k$ represents the number of blocks.

Expanding from digital electronic computing, Oguz et al. \cite{oguz2023forward} utilized the recently proposed contrastive loss-based local learning scheme, forward-forward algorithm (FFA) \cite{hinton2022forward}, to train optical neural networks. This study demonstrates experimentally that multimodal nonlinear optics can significantly improve the performance of multilayer NN architectures without creating additional computational overhead or extensive characterization experiments. Furthermore, Momeni et al. \cite{momeni2023backpropagation} proposed physical local learning (PhyLL). Unlike FFA, PhyLL leverages cosine similarity between two forward passes -- one for positive data and one for negative data -- eliminating the need for layernorm operation, which can be challenging to implement physically. This approach was evaluated experimentally across three PNNs (Acoustics, Microwave, and Optics), allowing for supervised and unsupervised training without detailed knowledge of the nonlinear physical layer’s properties. 
Another promising avenue is to adapt methods from self-supervised learning \cite{Federici2020Learning,balestriero2023cookbook}, which may also be well-suited to training PNNs in a way that avoids gradient communication between layers \cite{laydevant2023benefits}.
A challenge for training arbitrary physical systems with PhyLL is that it uses knowledge of the behavior of each individual layer so that an estimation of the gradient for each layer can be computed and used when updating parameters \cite{momeni2023phyff}.   More recently, Zhao et al. \cite{zhao2023high} used a Monte-Carlo gradient-estimation algorithm to compute the required gradient for in-situ updating the parameters of PNNs. 

While local learning has great potential to scale up in terms of hardware, it remains far from clear whether these methods can, at any scale above small laboratory demonstrations, reproduce the performance of backpropagation. While exactly matching backpropagation is not necessary (especially given the potential for radically improved efficiency), going forward, such guaranteed high-dimensional scaling is an essential requirement for physical local learning techniques.

\begin{figure}
    \centering
    \includegraphics[width=1\textwidth]{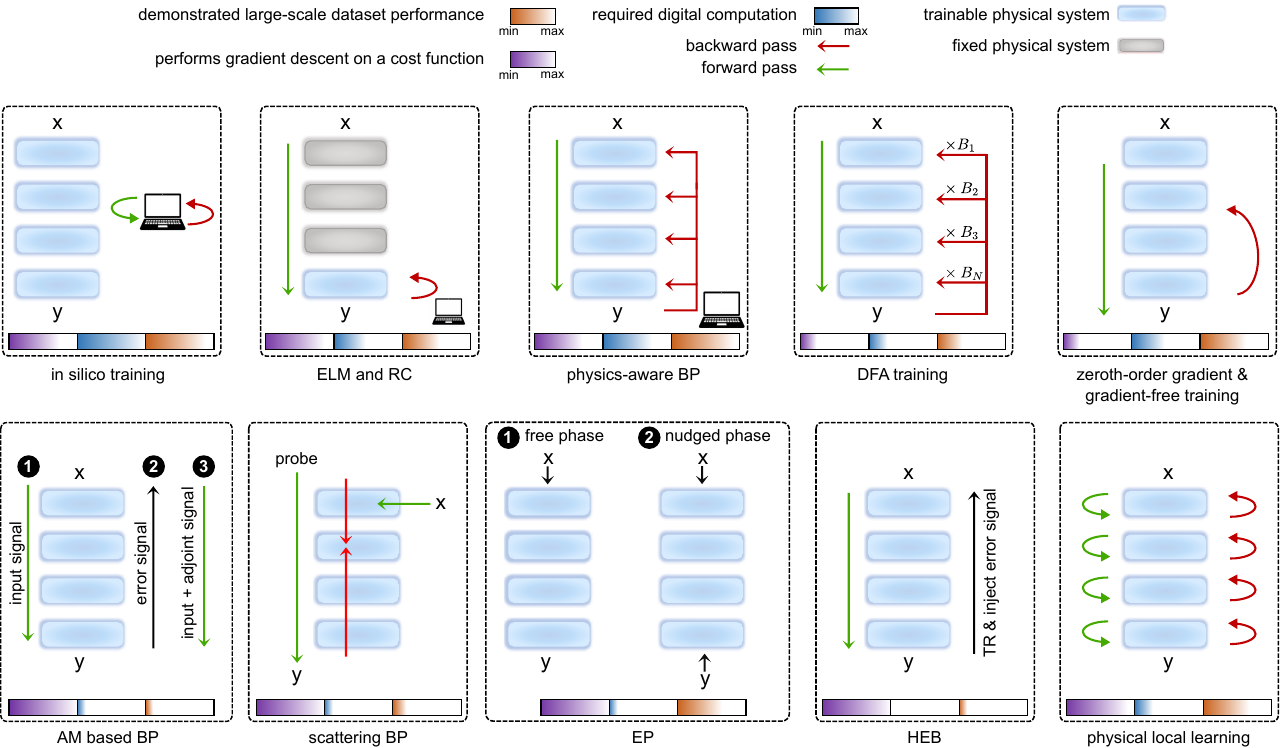}
    \caption{\textbf{Training methods for PNNs.} ELM: Extreme Learning Machine, RC: Reservoir Computing, DFA: Direct Feedback Alignment, EP: Equilibrium Propagation, HEB: Hamiltonian Echo Backpropagation. For a more detailed comparison, refer to Table~1.}
    \label{fig:enter-label}
\end{figure}

\subsection*{Zeroth-Order Gradient and Gradient-free Training}
To eliminate the need for detailed knowledge of the physical system whatsoever, model-free, "black-box," or gradient-free training algorithms have been proposed.
However, fully-fledged implementations of them in hardware remain scarce. These methods are typically slow because the number of gradient updates scales linearly with the number of learnable parameters in the network, posing a significant challenge for scaling up. These algorithms can be split into two broad categories. On the one hand, perturbative methods estimate the gradient by sampling a target function to optimise (i.e. loss function) at different coordinates (i.e. weights), and after estimating a gradient, weights are optimised via traditional gradient descent. The finite difference method is the simplest way to estimate a gradient by sequentially perturbing each weight and computing its corresponding gradient. More advanced zeroth-order methods have been developed, such as the Simultaneous Perturbation Stochastic Approximation (SPSA) algorithm \cite{spall1992multivariate}, which perturbs all weights simultaneously. Variants of this algorithm have been used in electronics \cite{wang2019memristor} and optics \cite{bandyopadhyay2022single,gu2020flops}. Although simple, these algorithms have achieved good performance in the context of on-chip in-situ training \cite{gu2021l2ight,mccaughan2023multiplexed}.
On the other hand, the second class of gradient-free methods consists of population-based sampling. These include popular classes of algorithms such as genetic algorithms (GA), surrogate optimization~(SO), evolutionary strategies (ES), swarm optimization and reinforcement learning (RL) algorithms. These are not directly concerned with achieving an approximation of the gradient but rather iteratively generating better candidate solutions to an optimization problem either according to heuristic criteria in the case of GA, ES, and swarm-type algorithms or according to an iteratively improved candidate generation policy in the case of RL. Evolutionary strategies such as the CMA-ES \cite{hansen2016cma}, metamodel-based optimization\cite{oguz_programming_2024}, as well as RL have been used to train optical neural networks in\cite{skalli2022computational,bueno2018reinforcement,lupo2023deep,kanno2020adaptive}.

\subsection*{Gradient-Descent Training via Physical Dynamics}

Gradient descent optimization is the workhorse of state-of-the-art machine learning systems. We present four physical training methods that achieve gradient descent without needing a digital twin. Such methods can potentially lead to energy gains of 4 orders of magnitude compared to GPU-based neural network training \cite{yi2023activity}.

The first approach aims at mapping traditional neural networks and BP onto analog hardware. The central insight is that the matrix-vector multiplications required in the forward pass (inference) and backward pass (training) can be implemented using linear reciprocal physical systems \cite{Burm2015trainable}, e.g. as a wave propagation through a linear medium in photonics systems \cite{Burm2015trainable}, or using memristor crossbar arrays in electrical circuits \cite{lebleici2017neuromorphic}. In photonic systems, gradients can be efficiently calculated in situ by backpropagating the adjoint electromagnetic field and by interfering its (forward-propagating) time-reversed copy with the original forward field \cite{hughes2018training, pai2023experimentally}. This implementation requires bidirectional propagation of complex amplitude waves (complex field generation modules are required \cite{zhou2020situ, Miller2017setting}) and the (ideally lossless) measurement of the wave intensity in the network (to this aim in photonics almost transparent detectors could be used \cite{Morichetti2014noninvasive}). This approach requires the network to be lossless, or at least with uniform loss across the network to preserve unitarity. Most often, these photonics and electrical implementations of BP are mixed-signal, executing the nonlinear activation function and its derivative in the digital domain \cite{hughes2018training, pai2023experimentally}. In order to avoid analog-digital and digital-analog converters, and thus further improve energy efficiency, other methods have been proposed to extend physical backpropagation to lossy networks \cite{zhou2020situ} and to nonlinear activation layers \cite{Guo2021backpropagation, Spall2023training}. However, non-idealities of physical non-linearities will cause deviations between calculated and true gradients.

The second approach is based on nonlinear computation via linear wave scattering \cite{Wanjura2023fully, Yildirim2023nonlinear, Xia2023deep} (also used in \cite{momeni2023backpropagation}). Here, input data are encoded in tuneable, physical parameters, such as the frequencies of optical resonators, while other parameters are optimized during training; and the scattering response serves as output of the neuromorphic system. As a significant advantage, all the gradients can be obtained through a minimal number of single-shot scattering experiments \cite{Wanjura2023fully} without the need for complete knowledge or control of the system. 

Equilibrium Propagation (EP) is the third approach \cite{scellier2017equilibrium}. EP applies in energy-based systems, i.e. systems in which physics tends to minimize an “energy” or Lyapunov function. The input is supplied as a boundary condition while physics drives the system to an energy minimum (equilibrium) to produce the response (output). In EP’s original formulation, the weights are updated by a local contrastive rule based on comparing two equilibrium states corresponding to two different boundary conditions. As a major advantage over other contrastive learning algorithms \cite{ackley1985learning,movellan1991contrastive,stern2021supervised}, EP calculates the weight gradients of arbitrary cost functions \cite{scellier2017equilibrium,laborieux2022holomorphic,scellier2023energy}. Examples of energy-based systems trainable by EP include continuous Hopfield networks \cite{scellier2017equilibrium}, nonlinear resistor networks \cite{kendall2020training}, Ising machines \cite{laydevant2023training} and coupled phase oscillators \cite{wang2024training}. Experimental realizations of the contrastive rule pose a challenge since it requires comparing network states under two different sets of boundary conditions. An EP-like contrastive scheme has been successfully implemented in memristor crossbar arrays \cite{yi2023activity}, and a binary version of EP implemented on D-Wave solved MNIST with software-equivalent accuracy \cite{laydevant2023training}. Both implementations resorted to external memory to store the states between the two phases. Another contrastive scheme called Coupled Learning (CL) \cite{stern2021supervised} has been experimentally demonstrated in elastic networks \cite{altman2023experimental}.

To realize energy gains, however, EP must be implemented in the lab without digital processing or external memory. Potentially scalable laboratory prototypes have been developed of such electrical linear \cite{dillavou2022demonstration} as well as nonlinear \cite{dillavou2023machine} resistor networks; these implement CL by coupling two copies of the network\cite{dillavou2022demonstration,dillavou2023machine}. Further energy gains can be realized by including the energy as an additional term in the cost function \cite{stern2024training}. Other proposed solutions to implementing the contrastive rule resort to encoding the two states in different physical domains \cite{anisetti2024frequency}, using integral feedback \cite{falk2023contrastive}, working in the complex domain \cite{laborieux2022holomorphic} or using dynamics of spiking networks \cite{martin2021eqspike}. Another conceptual advance is a non-contrastive version of EP where the weights are physically updated through physical equilibration \cite{scellier2022agnostic}. The potential of EP has further been highlighted in simulations by training energy-based convolutional networks on a down-sampled version of the ImageNet dataset \cite{laborieux2022holomorphic}.

Hamiltonian Echo Backpropagation (HEB) is the fourth approach \cite{lopez2023self}. On top of extracting the weight gradients, HEB directly produces the correct weight updates using physical dynamics, without any feedback. HEB applies in time-reversal invariant Hamiltonian systems where dissipation is (ideally) absent. Another crucial ingredient is a "time-reversal operation", e.g. phase conjugation in nonlinear optics experiments. During training, in the forward pass, a signal wave and a trainable-parameter wave travel jointly through a nonlinear medium where they interact. An error signal is superimposed on the signal wave, and a time-reversal operation sends both waves back through the medium. At the end of this backward pass, the trainable wave has been automatically updated in the direction of the cost-function gradient.

%\begin{figure}
%    \centering
%    \includegraphics[width=.55\textwidth]{figures/EPandHEB_v5.pdf}
 %   \caption{\textbf{a}~Equilibrium propagation (EP) allows to compute gradients by comparing correlation functions computed in two different equilibria: in the free phase, the input units are fixed and all other units evolve to an equilibrium. In the nudged phase, an additional potential (cost function) nudges the output units towards their target and all other units evolve towards a new equilibrium. The difference of correlation functions yields the gradients for the weight updates.
 %   \textbf{b}~Illustration of the working principle behind Hamiltonian echo backpropagation (HEB). During the forward pass, a charged ball $\Psi$ (representing the input) follows a trajectory that is deflected by a second ball $\Theta$ (representing the weight parameters) which can move freely along a rail and moves due to the interaction with ball $\Psi$. In the second stage, a time-reversal operation is performed and an error signal injected, i.e. the velocities of both balls are reversed and the position of ball $\Psi$ is corrected according to the deviation from its desired outcome. During the backward pass, the ball $\Psi$ almost exactly retraces its trajectory and the ball $\Theta$ moves again due to the interaction with the other ball. Any remaining velocity of the ball $\Theta$ is dissipated leading to a finite displacement representing the parameter update.}
  %  \label{fig:enter-label}
%\end{figure}

\subsection*{Continual Learning}
Neural networks are usually trained ``off-line'' on a fixed dataset, and then deployed for inference, without further training. Continual learning aims to enable neural networks to learn from non-stationary streams of data incrementally. It is not a trivial problem: when trained on a new dataset, neural networks tend to lose their previously acquired capabilities by overwriting weights involved in representing the old learning. This problem is known as ``catastrophic forgetting''\cite{van2022three}. To address this, research efforts have focused on freezing parts of the network weights while simultaneously growing other parts of the network to extend the learning ability. For example, in class-incremental learning (CIL), the network must incrementally learn to distinguish novel classes without forgetting the previously observed classes\cite{van2022three}. Very recently, few-shot CIL algorithms have been proposed, in which the continual learning of novel classes is done with only a few (e.g. 5) data samples, which is even more challenging\cite{zhang2021few,hersche2022constrained}.  

PNNs are excellent choices for implementing few-shot CIL algorithms due to their capacity for continual expansion, allowing them to accommodate new classes effectively.  For example, PNNs implemented with arrays of memristive devices can be incrementally expanded by programming previously unused devices by applying suitable electrical pulses, which will retain information about the novel classes in a non-volatile way. Such an implementation was realized based on an enhanced memory-augmented neural network comprising a dynamically growing explicit memory implemented with a phase-change memory (PCM) array\cite{karunaratne2022memory}. The novel classes were learned and stored incrementally in the explicit memory by exploiting the in-situ progressive crystallization of PCM devices, and an in-memory similarity search was performed during inference in the PCM array to classify unseen examples~\cite{karunaratne2022memory}. 

Another opportunity to implement continual learning in PNNs is offered by  Ising or XY machines, which are specialized hardware engineered to solve optimization problems. They exploit their inherent ability to discover low-energy states in a spin system naturally\cite{Stroev2023Analog}. The architecture of such machines, often realized through optical, light-matter, or quantum systems, inherently supports parallel computations, making them suitable for continual dynamic learning. To enable continual learning, the XY or Ising machine's architecture can be modified to dynamically adjust the interactions between spins \cite{mohseni2022ising,stroev2021neural}. Such adaptability can be facilitated by developing algorithms that incrementally update the Hamiltonian—the system's underlying energy function—reflecting the continual integration of new information and the retention of existing patterns while avoiding catastrophic forgetting.

Further work is nonetheless required to implement an entire CIL architecture in PNN hardware to demonstrate fully end-to-end continual learning more efficiently than in a traditional digital architecture.

\begin{figure}
    \centering
    \includegraphics[width=.75\textwidth]{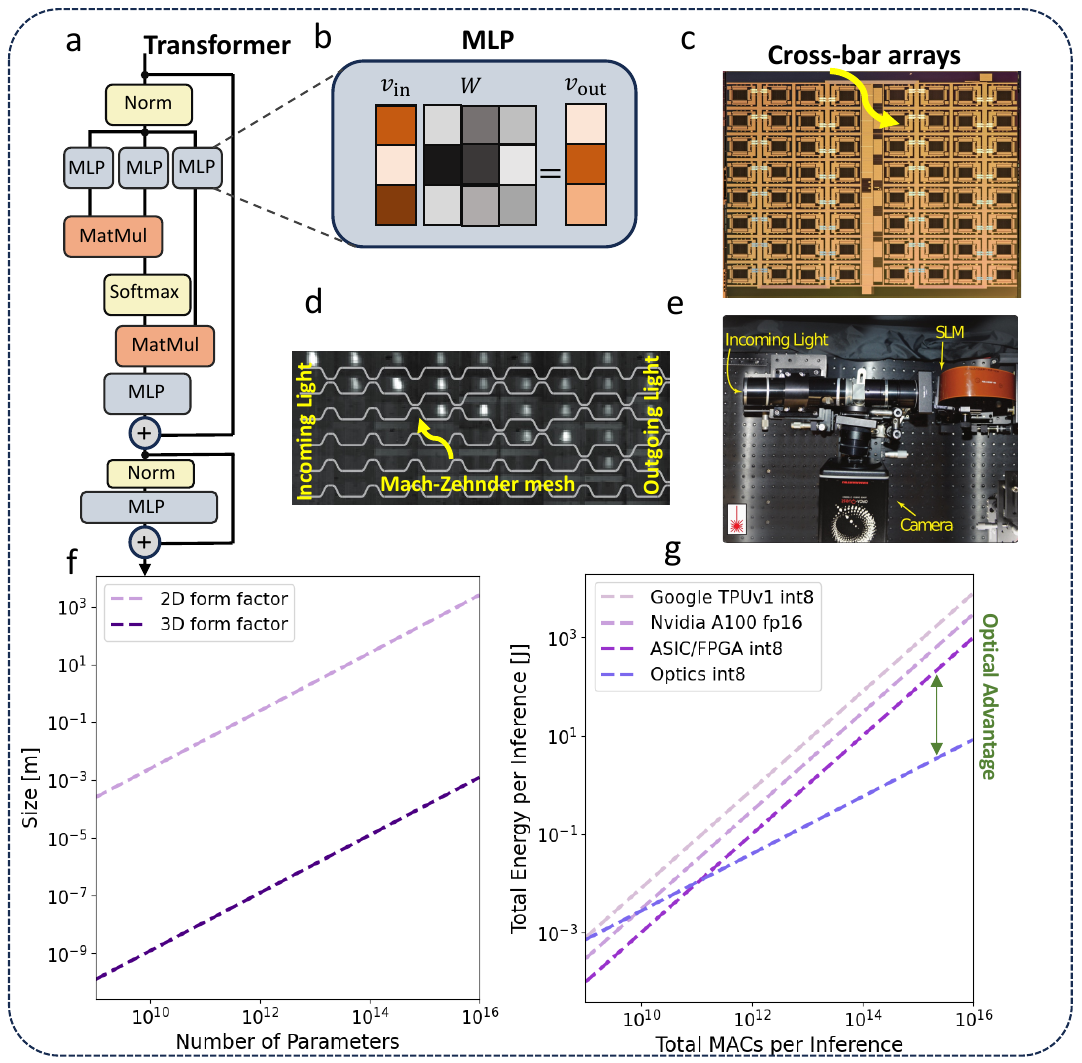}
    \caption{{\bf Analog large models} (a) 
    The building block of the mainstream large language models is the transformer architecture \cite{vaswani2017attention}, whose main building blocks are the attention, multilayer perceptron (MLP) layers, softmax operation and dynamic matrix-vector multiplication (MatMul). The attention layer requires a causal pairwise computations between the elements in the sequence, resulting in a quadratic increase in computational complexity with respect to sequence length, affecting both time and energy overhead, especially as models process longer context lengths. The MLP layer includes very large weight matrices that also impose a large computational overhead; (b) MLP is the architecture of vector-matrix multiplication also known as a fully connected layer. The MLP can be experimentally realized on a number of technologies such as (c) crossbar arrays \cite{le202364}; (d) Mach-Zehnder Interferometer meshes \cite{pai2023experimentally}; (e) free-space multipliers \cite{anderson2023optical}; (f) size scaling of two- and three-dimensional analog models with increasing model parameters computed at wavelength=500 nm with scalings $\lambda^{2/3}$; (g) Energy scaling advantage of analog optical matrix-vector multiplication compared to digital electronics, for Transformer models. Data were obtained from \cite{anderson2023optical}.}
    \label{fig:mvm}
\end{figure}

%\fbox{\begin{minipage}{42em}

%\section*{Box2: Key conclusions and recommendations:}

%\textbf{Conclusions}

%(1)

%(2)

%(3)

%\textbf{Recommendations}

%(1)

%(2)

%(3)

%\end{minipage}}

\section*{Towards Implementation of Analog Efficient Large Models}
In the next two sections, we first explore large AI models and briefly review strategies to improve their efficiency from a digital standpoint, both in terms of hardware and software innovations. Following that, we consider the potential of "analog" large models, i.e., large AI models implemented by analog PNN hardware. While PNNs are currently far from competitive with digital approaches (which are themselves progressing rapidly), PNNs possess unique physical properties that may make them an important route for scaling AI models beyond the practical limits of conventional digital implementations. 

\subsection*{Efficient Training and Fine-tuning of Large models}

In recent years, we have witnessed remarkable improvements in the capabilities of neural networks, especially those designed in rich data setting including language understanding such as GPT-2 (1.2B)\cite{radford2019language}, GPT-3 (175B)\cite{mann2020language}, LLaMA (65B)\cite{touvron2023llama}, PaLM (540B)\cite{chowdhery2023palm}, GPT-4\cite{achiam2023gpt}, Gemini\cite{geminiteam2024gemini}, vision-language understanding such as CLIP\cite{radford2021learning} and LLaVA \cite{liu2024visual}, scientific reasoning including MatterGen\cite{zeni2023mattergen} and AlphaFold\cite{jumper2021highly}, and climate prediction \cite{andrychowicz2306deep}. Demonstration of emerging abilities in these large models, mainly based on the attention architecture in Fig. \ref{fig:mvm} a, is driven by "scaling laws"—a trend indicating that a model's ability to generalize improves in a predictable, log-linear manner as a function of number of parameters, data examples, or the amount of compute used to model and train large models \cite{radford2019language}. Training frontier LLMs with hundreds of billions of parameters is a prohibitively expensive task that requires weeks (or even months) of optimization on thousands of accelerators (e.g. Nvidia GPUs) and trillions of tokens of data. The inference cost of LLMs is also high due to their large memory footprint, which requires model partitioning across many accelerators (causing I/O overhead), as well as the large number of floating point operations across the model.

To address these challenges, the machine learning research community has been actively exploring strategies to enhance model efficiency without unduly compromising model accuracy. These methods include architectural changes such as moving away from attention-based architectures, quantization of model parameters, fine-tuning methods to avoid costly retraining, and efficient implementation on digital hardware.

One such effort has been on designing sublinear attention mechanisms. Sub-quadratic attention models replace full attention by taking advantage of attention properties such as low-rankness \cite{xiong2021nystromformer} and sparsity\cite{child2019generating}. In these models, the softmax operation is replaced by linear\cite{katharopoulos2020transformers}, low-degree polynomials\cite{zhang2024hedgehog} or random feature maps\cite{peng2021random}. An orthogonal approach is replacing the attention with recurrent models \cite{gu2023mamba}. Another effective approach for performance efficiency is quantization, in which the model weights are transformed to a lower precision (e.g. from 16 bits to 4 bits)\cite{frantar2022optq,lin2023awq}. Quantization is commonly performed as a post-training step, however, recent methods on quantization-aware training of LLMs show that models with as low as 1.58-bit (ternary parameters) can match the performance of full-precision transformers, saving the energy consumption by tenfold \cite{ma2024era,wang2023bitnet}. In addition to (pre-)training, fine-tuning is an important phase in LLM optimization process. During fine-tuning, a pre-trained model is adapted to a specific task, using a relatively small dataset (often in the order of thousands of examples or less). Fine-tuning models are commonly done through supervised or reinforcement learning\cite{christiano2017deep, rafailov2024direct} and would require updating the entire model parameters. All techniques applied to efficient training are also applicable to fine-tuning. Moreover, there are techniques specifically developed for fine-tuning, such as LoRA\cite{hu2021lora} and Adapters\cite{houlsby2019parameter, lin2020exploring}, in which updating only a small subset of the parameters (often those that are newly added) is sufficient to adapt the model for downstream tasks. Digital hardware-aware implementations of the exact attention mechanism have also been shown to be very effective. For example, FlashAttention\cite{dao2022flashattention} uses tiling and recomputation to reduce the memory bandwidth overhead, Hydragen\cite{juravsky2024hydragen} splits the sequence across the prompt and suffix and batches attention queries over the shared prompt sequences to increase throughput. vLLM \cite{kwon2023efficient} avoids redundant storage of the prefix keys and reduces the cost of redundant reads.

\subsection*{Analog Large Models}

AI models are now large, and they are getting larger. As the previous section summarized, many aspects of these models are likely to change, and digital software and hardware innovations will help reduce their energy costs in both training and inference. Could PNNs make this scaling easier? Could they allow scaling beyond what is economically feasible for digital computers?

As a first consideration, large AI models are literally physically large. In optics, for example, parameters are often encoded in the pixels of spatial light modulators (SLMs). A 1-quadrillion ($10^{15}$) model would require a total area of SLMs of roughly 16 m$^2$, comparable to the cross-section of an elephant. Does this mean that large optical PNNs have no hope? No -- in fact for computations of this scale, a large footprint is inevitable with any hardware. However, it does suggest that PNN researchers (across all hardware) will need to propose architectures designed not just for laboratory demos, but for extreme scalability. One initial step are the projections of Anderson \textit{et al.}, which examine the costs of implementing Transformer models using optical hardware \cite{anderson2023optical}. This work confirms that, even when splitting a model across many reasonably-sized optical processing units and accounting for the costs of loading parameters and inputs from memory, plausible optoelectronic hardware could be roughly 100x more efficient than 2023's state-of-the-art digital electronics at implementing current LLMs (about $10^{15}$ MACs/inference). More importantly, this advantage could grow to $10^4$ or even $10^5$ for larger models with $10^{19}$ MACs/inference. 

This highlights perhaps the most important scaling consideration for potential future large-scale PNN AI systems: If PNN hardware is designed properly, its different underlying physics may allow it to exhibit different energy scaling behavior than digital electronics. This means that, given sufficiently large model scale, PNN implementations may offer better efficiency compared to digital systems, even despite the many overhead costs of analog hardware, such as digital-to-analog conversion costs. In optics for example, assuming fixed output precision, the optical-energy cost per operation (scalar MAC) when performing a dot product scales as $1/N$, where $N$ is the length of the vector \cite{shen2017deep,hamerly2019large,nahmias2019photonic,wang2022optical}. In digital electronics, the cost per operation is normally fixed ($\sim 1$). This optical dot product energy scaling advantage ($\sim 1/N$ versus $\sim 1$) may transfer to a similar scaling advantage for AI model inference, since most models consist primarily of dot products. However, this transfer is not guaranteed: the algorithm and data, design of the electronic modulation and memory access, and the optical hardware itself can limit the scaling advantage \cite{anderson2023optical,tait2022quantifying}. 

Finally, scaling is not only about hardware. While Transformers are a breakthrough, they are also merely the latest algorithm to rise to prominence because of synergy with scalable hardware \cite{laydevant2024hardware,hooker2021hardware}. As we look to ultra-scale PNNs, it may be shortsighted to focus on current algorithms -- rather, we need new synergistic combinations of hardware and software. Given the inertia of infrastructure and the rapid development of efficient digital large model implementations, commercially viable PNNs will need to provide energy efficiency thousands if not millions of times greater than digital electronic. To do this will require designing physical computers that consider the challenge of scale holistically, hardware and software together, and that make efficiently exploiting physical compute their leading objective \cite{laydevant2024hardware}. 

All the above considers primarily inference of large models, which is the most urgent and accessible opportunity for PNNs in the large-model space. Using PNNs to accelerate training of such models is similarly promising. As this article's earlier sections suggest, the training phase may allow additional physical phenomena to be gainfully exploited, e.g., to realize scalable local learning based on analog physical processes. This means that physics-driven learning of PNN models may exhibit additional scaling advantages (in energy, size, speed, etc.) that extend even beyond the inference energy scaling advantages noted above.

\section*{Emerging PNN Technologies}

In the PNN context, quantum, probabilistic, photonic,  light-matter and hybrid computing represent promising developments \cite{Stroev2023Analog}. 
%Each offers distinct computational and physical mechanisms that align with the core objectives of PNNs - efficiency, scalability, and reduced energy consumption. 
%Quantum computers, particularly quantum annealers, apply quantum mechanics principles such as quantum tunnelling and superposition to address optimization problems critical to NN training \cite{cerezo2022challenges}.
Quantum computers can exploit features of quantum mechanics such as superposition in a way that might allow them to address optimization problems critical to NN training \cite{cerezo2022challenges}.
%By exploiting quantum tunnelling and superposition, these systems can navigate complex energy landscapes more efficiently than classical approaches, offering a promising pathway to accelerate PNN training while potentially minimizing energy use.  Superposition allows quantum bits (qubits) to exist in multiple states simultaneously, offering a multidimensional computational space. Entanglement, where the state of one qubit instantaneously influences another regardless of distance, can introduce unprecedented parallelism and connectivity in neural networks. 

\begin{figure}
    \centering
    \includegraphics[width=1\textwidth]{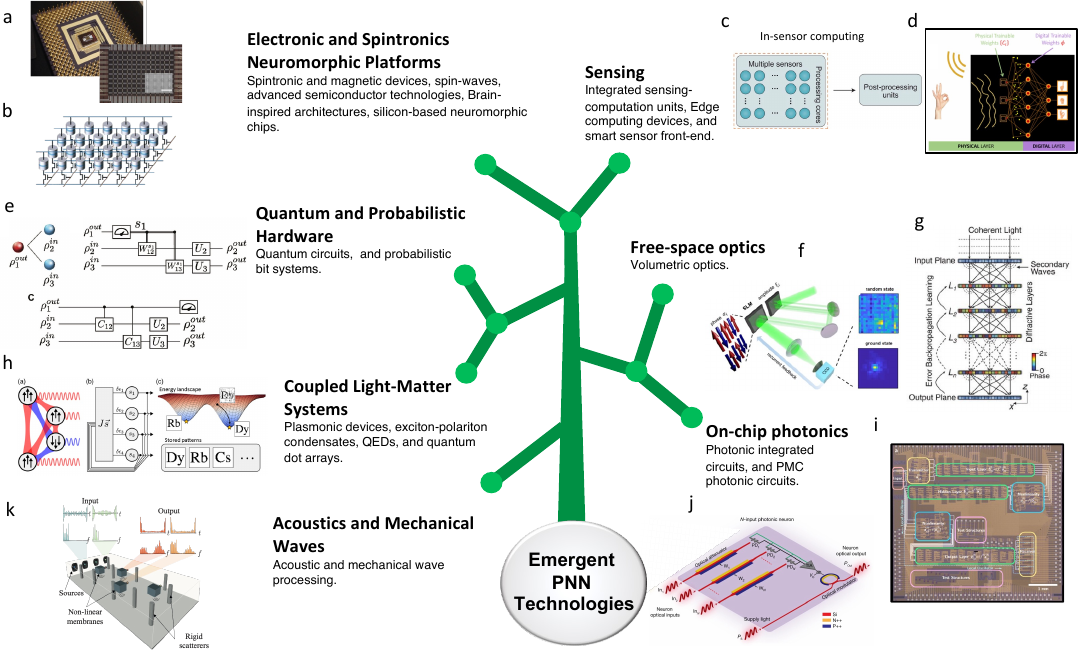}
    \caption{{\bf Emergent technologies.} (a) Optical microscopic image of a memristor crossbar array integrated on the memristor/CMOS chip, from \cite{cai2019fully}; (b) crossbar array of magnetic tunnel junctions for high-density storage and memory retrieval, from \cite{grollier2020neuromorphic}; (c) the schematics of  in-sensor computing architecture, from \cite{zhou2020near}; (d) an operational principle of learned-sensing intelligent meta-imagers, from \cite{saigre2022intelligent}; (e) an illustration of soft quantum neurons in the quantum circuit model, from \cite{zhou2023quantum}; (f) the schematics of spatial photonic Ising machine, from \cite{pierangeli2020noise}; (g) diffraction optical NNs consisting of multiple transmissive or reflective layers, where each point on a given layer acts as a neuron, with a complex-valued transmission or reflection coefficient, from \cite{lin2018all}; (h) superradiance in confocal cavity QED for high-density storage and memory retrieval, from \cite{marsh2021enhancing}; (i) an image of a  PIC with shown signal paths (white) and the local oscillator paths (blue), from \cite{bandyopadhyay2022single}; (j) the structure of an N-input photonic neuron with weights of input signals  changed using optical PIN attenuators and summed up using photodetectors, from \cite{ashtiani2022chip}; (k) acoustic data transformer where input data are encoded into the intensity of sound waves at different frequencies that propagate through a random set of membranes, from \cite{momeni2023backpropagation,momeni2023physics}.
    }
    \label{fig:emerging}
\end{figure}

However, the practicality of applying these quantum advantages is tempered by the limitations of current Noisy Intermediate-Scale Quantum systems, which have limited qubits and large computational error rates \cite{kashif2024hqnet}. Specific quantum algorithms and quantum neural network frameworks are being devised to operate within these constraints, for instance, using `soft quantum neurons' \cite{zhou2023quantum}, quantum circuit Born machines, quantum generative adversarial networks \cite{tian2023recent}, and variational quantum algorithms \cite{cerezo2021variational} can potentially outperform classical models in generating new samples and learning data distributions.

Probabilistic hardware systems, which rely on stochastic elements, sometimes referred to as probabilistic bits (p-bits) occupy an intermediate step between quantum and classical deterministic PNNs. They are naturally suited to training deep generative models, specifically deep Boltzmann Machines (DBMs) \cite{niazi2023training}, but can also be trained to act as stochastic neural networks for deterministic classification tasks \cite{ma2023quantum}.

%A Probabilistic Ising machine was employed to train deep generative models, specifically deep Boltzmann Machines (DBMs) \cite{niazi2023training}. These machines use sparse, asynchronous, and massively parallel configurations to enhance training efficiency and performance. The p-bits in these machines simulate the stochastic neurons in Boltzmann networks, allowing the hardware to generate samples from the Boltzmann distribution at speeds significantly higher than traditional  CPUs or GPUs. Besides probabilistic machines, various atomic spin systems can emulate a Boltzmann machine, for instance, using the orbital dynamics of individual cobalt atoms on black phosphorus. The synaptic weights autonomously reorganize in response to external electrical stimuli. This demonstrates a self-adaptive architecture that directly paves the way for autonomous learning in atomic-scale machine learning hardware \cite{kiraly2021atomic}.
Photonic-based optimisers   such as Spatial Photonic Ising Machines (SPIMs) that use spatial light modulation to emulate Ising problems introduce another opportunity for PNNs \cite{pierangeli2020noise}. The properties of light, such as  spatial parallelism achievable in optics and the dissipationless dynamics of light propagation that enables computation, offer significant  advantages over electronic systems \cite{mcmahon2023physics}. 

Light-matter systems that couple photons to matter particles efficiently \cite{keeling2011exciton, carusotto2013quantum} form the basis of gain-based computing that could lead to new methodologies in PNNs by encoding optimization problems within driven-dissipative systems' gain and loss rates ~\cite{Berloff2017,marsh2021enhancing}. For instance, polaritonic systems with their strong noise-controlling nonlinearities were proposed to lead to all-optical platforms for implementing diffusion models \cite{sohl2015deep,johnston2024macroscopic}.  

  A limitation of the stand-alone PNNs is the overhead from encoding inputs into and reading outputs from the physical domain.
  Integrating wave-based computing and PNNs at a sensor front-end naturally avoids such overhead. Intelligent sensors combine sensing and over-the-air computing capabilities to pre-select task-relevant information during data acquisition, yielding substantial improvements in latency and other metrics\cite{saigre2022intelligent}. Intelligent sensors are to date conceived by training via error backpropagation the entire sensing pipeline end-to-end concerning a specific task, including both analog data acquisition and digital post-processing\cite{horstmeyer2017convolutional,sitzmann2018end}. In other words, training intelligent sensors to date requires a differentiable digital model; alternative training paradigms remain unexplored. The hallmark feature of an intelligent sensor is that its data acquisition is task-aware thanks to the end-to-end training\cite{saigre2022intelligent}. Endowing the physical layer with programmability and/or non-linearity enables task-reconfigurable smart sensors and/or increases the complexity of mathematical operations that can be performed during data acquisition\cite{del2020learned,zhou2021large,liu2022programmable,wang2023image}. These paradigms are emerging across scales and wave phenomena, and context-aware next-generation wireless networks may already leverage dynamic metasurface antennas for integrated sensing, computing and communications\cite{del2020learned,you2021towards,rajabalipanah2019addition,tahmasebi2022parallel,rajabalipanah2022parallel}. Other applications include privacy-preserving cameras\cite{bai2022image}, non-destructive testing\cite{li2023rapid} and noise-adaptive smart computational imagers\cite{qian2022noise}. A trend toward further integration of wave-based computing with sensing, communications and data storage is clearly emerging\cite{rios2019memory,sludds2022delocalized,rahman2023learning}.

The programming of all-photonic routers in networks also bears significant similarities to the training of PNNs, and these fields can benefit from one another. Irrespective of the detailed implementation\cite{lopez2020auto,sol2023reflectionless,mengu2023diffractive,dinc2024multicasting,seyedinnavadeh2024determining}, programmable all-photonic routers are (usually linear) input-output systems with a multitude of tunable degrees of freedom. The latter must be reconfigured during runtime to realize different routing functionalities (i.e., implement different input-output relations). A further important constraint is that reflections back into the input channels should be suppressed to avoid reflected-power echoes in the network\cite{sol2023reflectionless}. Besides various well-established global optimization techniques, ideas for progressively configuring specific hardware architectures purely based on local feedback loops are emerging\cite{seyedinnavadeh2024determining,miller2013self}, even for only partially coherent light\cite{roques2024measuring}.

Integrating these advanced computational paradigms into PNNs requires addressing several challenges, including adapting learning algorithms to leverage quantum and photonic, wave-based or gain-based processes, managing noise and error rates in quantum systems, and the scalability of architectures. The development of hybrid systems, which combine quantum or photonic processing units with classical computational elements, might offer practical pathways to use the advantages of these technologies while mitigating their current limitations. Aligning the unique properties of these physical systems with the goals of PNNs can pave the way for the next generation of intelligent systems characterized by unprecedented levels of speed, efficiency, and scalability.

Finally, beyond human-made devices, the dynamics of biological systems may offer enticing opportunities to implement new generations of PNNs, e.g., using octopus-inspired soft robotic arms for reservoir computing\cite{nakajima2013soft,momeni2022learning}. Biological contexts may require rethinking the training process of PNNs to enable their compatibility with the vulnerability of underlying biological systems.

\section*{Outlook}
PNNs may ultimately be found from the data center to the edge, from powering large generative models, through to aiding in classification in smart sensors. In all cases, they will need to be trained, but depending on the application, the constraints on training may be different. For example, large models on servers might only require updates every few months and be able to use a lot of energy to be trained, whereas some models at the edge might need to be adaptive on a timescale of hours or even minutes, and be severely limited in their power budget for retraining. The diversity of PNNs and use cases suggests that the major open challenge for the field is not to find what the single best training method is, but rather what the best training method for each situation is, and what the tradeoffs between different methods are.

An ideal training method would:

1)	be model-free, not requiring the training procedure to have access to a mathematical description of the behavior of the PNN hardware, and would rely on as few assumptions about the behavior and structure of the PNN hardware as possible, allowing PNN designers to optimize the hardware for speed and energy benefits in inference;

2)	give speed and energy advantages in the training time and energy cost versus training a conventional artificial neural network to perform the same task with the same accuracy, using the same training data – to achieve such benefits in training, the method should leverage the benefits of the PNN hardware itself in training rather than relying heavily on digital-electronic hardware during training, and the training method should be able to exploit the full expressivity of the PNN hardware;

3)	be robust to hardware copy-to-copy variations, drift, and noise – and if not fully robust, then at least able to cheaply compensate for these imperfections.

None of the known training methods for PNNs simultaneously satisfy all of these properties, or even perfectly satisfy any one of them. However, the past few years have seen many different training methods be developed that push the boundaries of tradeoffs within this space of properties, and we anticipate further advances that will lead to methods that are simultaneously more general, more efficient, and more robust, enabling practical and widespread use of PNNs.

\textbf{Acknowledgements}
We thank J\'er\'emie Laydevant, Martin Stein, and Mandar Sohoni for helpful feedback on a draft of this manuscript.
R.F. and A.M. acknowledge funding from the Swiss National Science Foundation (SNSF) under the Eccellenza award 181232. R.F. and P.d.H. acknowledge funding from the ANR-SNSF MINT project 212577 entitled ``Ultra-compact non-linear metamaterial wave processors for analog deep learning''. P.L.M. acknowledges funding from the National Science Foundation (award CCF-1918549) and a David and Lucile Packard Foundation Fellowship. N.G.B.~acknowledges the support from the HORIZON EIC-2022-PATHFINDERCHALLENGES-01 HEISINGBERG project 101114978 and Weizmann-UK Make Connection grant 142568. S.G. is a member of the institut Universitaire de France. T.O., L.G.W, and P.L.M. thank NTT Research for their financial and technical support.

%\textbf{Author Contributions}

\textbf{Competing interests} T.O., L.G.W. and P.L.M. are listed as inventors on a US provisional patent application (number 63/178,318) on physical neural networks and physics-aware training.



\newpage

\fontsize{7}{9}\selectfont  
\centering  
  
\begin{longtable}{|p{2.5cm}|p{3cm}|p{2cm}|p{2.5cm}|p{2cm}|p{2.5cm}|}  
\caption{Table for comparing different algorithms: $N$ is the number of parameters, $M$ is the number of neurons, and $T_0$ is time to convergence for backpropagation (BP).} \\  
\hline  
\textbf{Algorithm} & \textbf{Comments} & \textbf{Memory (Measurements)} & \textbf{Expected wall clock time to convergence} & \textbf{Updatable physical or digital parameters} & \textbf{Digital model (or simulation) required} \\  
\hline  
\endfirsthead  
% Repeat the header in case the table spans multiple pages  
\caption{Table for comparing different algorithms (continued).} \\  
\hline  
\textbf{Algorithm} & \textbf{Comments} & \textbf{Memory (Measurements)} & \textbf{Expected wall clock time to convergence} & \textbf{Updatable physical or digital parameters} & \textbf{Digital model required (Digital model accuracy)} \\  
\endhead  
\textbf{ELMs and RC} & Solution is found after one matrix inversion & $O(M)$ & Time to perform matrix inversion & Last digital linear layer & No  \\  
\hline  
\multicolumn{6}{|l|}{\textbf{\textcolor{blue}{Backpropagation-based methods}} }\\  
\hline  
\textbf{In-silico BP} &  & $O(M)$ & $O(T_0)$ & All digital parameters + physical parameters simulated in digital model & Yes \\  
\hline  
\textbf{Physics-aware BP} & Reduces constraints on model faithfulness & $O(M)$ & $O(T_0)$ & All digital parameters + physical parameters simulated in digital model & Yes \\  
\hline  
\textbf{Direct Feedback Alignment (DFA) training} &  & $O(M)$ & $>O(T_0)$ & Matrix elements of a digital or physical matrix-vector multiplier & Only for nonlinear
activation function
and its derivative \\  
\hline  
\textbf{Physical local learning} & Need the knowledge of each individual layer for estimating the gradients &  $O(M)$ &  $>O(T_0)$ (depends on the number of layers)& All controllable parameters &   No  \\  
\hline  
\multicolumn{6}{|l|}{\textbf{\textcolor{blue}{Zeroth-order/gradient-free methods}}} \\  
\hline  
\textbf{Finite difference stochastic approximation} &  & $O(N)$ & $O(N \cdot T_0)$ & All controllable parameters & No  \\  
\hline  
\textbf{Simultaneous perturbation stochastic approximation} &  & $O(N)$ & $O(N \cdot T_0)$  & All controllable parameters & No   \\  
\hline  
\textbf{Gradient-free training (GA/etc.)} && For Population-Based Methods (e.g., Genetic Algorithms, Evolution Strategies): $O(PN)$, where P is population size.   & $>>O(T_0)$ & All controllable parameters & No \\  
\hline      

\multicolumn{6}{|l|}{\textbf{\textcolor{blue}{Physical gradient computation/physical backpropagation}}} \\  
\hline 
\textbf{Adjoint Method (AM) based BP} & Reduces memory load & $O(1)$ & $O(T_0)$ & All controllable parameters & Only for nonlinear activation function and its derivative\\  
\hline  
\textbf{Scattering BP} & Nonlinear computation in physically linear systems; requires knowledge of form of Hamiltonian terms depending on tunable parameters; allows for batch processing using frequency multiplexing & $O(N_\mathrm{out} M)$ (with $N_\mathrm{out}$ the output dimension) & $O(T_0)$ & All controllable parameters & No  \\  
\hline  
\textbf{Equilibrium Propagation (EP)} & Applies to system converging to the minimum of an energy function. Requires knowledge of energy derivatives wrt trainable parameters. & Lazy implementation: $O(N)$, Advanced implementation: $O(M)$ & $O(1)$ & All controllable parameters & No  \\  
\hline  
\textbf{Hamiltonian Echo
Backpropagation (HEB) } & Applies to lossless systems with time-reversal operation & $O(0)$ & $O(1)$ & All controllable parameters &  No \\
\hline  
\end{longtable}

\newpage
\fontsize{9}{11}\selectfont
\bibliography{refs}


\end{document}